\documentclass[aps,preprint,showpacs,nofootinbib]{revtex4}
\usepackage{graphicx}
\usepackage{amssymb,amsmath,amsfonts,bm}
\usepackage{color}

\begin{document}

%\draft
%%%%%%%%%%%%%%%%%%%%% Title %%%%%%%%%%%%%%%%%%%

\title{
 Comments on differential cross section of $\bm\phi$-meson photoproduction at
 threshold }
%%%%%%%%%%%%%%%%%%%% Authors %%%%%%%%%%%%%%%%%%%%%
\author{A.~I.~Titov$^a$, T.~Nakano$^b$, S. Dat\'e$^c$, and Y. Ohashi$^c$}
 \affiliation{ $^a$Bogoliubov Laboratory of Theoretical Physics, JINR,
  Dubna 141980, Russia\\
 $^b$Research Center of Nuclear Physics, Osaka University,
 Ibaraki, Osaka 567-0047, Japan\\
 $^c$Japan Synchrotron Radiation Research Institute, SPring-8,
 1-1-1 Kouto Sayo-cho, Sayo-gun, Hyogo 679-5198, Japan}

%%%%%%%%%%%%%%%%%%%% Abstract %%%%%%%%%%%%%%%%%%%%%

\begin{abstract}
We show that the differential  cross section ${d\sigma}/{dt}$ of
$\gamma p\to \phi p$ reaction at the threshold is finite and its
value is crucial to the mechanism of the $\phi$ meson
photoproduction and for the models of $\phi N$ interaction.
\end{abstract}

 \pacs{13.88.+e, 13.60.Le, 14.20.Gk, 25.20.Lj}

 \maketitle

  Now it becomes clear that the $\phi$-meson photoproduction at
  low energies $E_\gamma\simeq 2-3$ GeV plays important role in
  understanding the non-perturbative  Pomeron-exchange
  dynamics and  the nature of $\phi N$ interaction.
 It was expected
 that in the diffractive region
 the dominant contribution comes from the Pomeron exchange, since
 trajectories associated with conventional meson exchanges
 are suppressed by the OZI-rule~\cite{NakanoToki}.
  The exception is the finite contribution
 of the pseudoscalar $\pi$,$\eta$-meson-exchange channel,
 but its properties are quite well understood~\cite{TL}.
 Therefore, the low-energy $\phi$-meson photoproduction may be
 used for studying the additional (exotic) processes.
 Candidates are the Regge  trajectories associated with
 a scalar and tensor mesons
 containing a large amount
 of strangeness~\cite{Will98,Laget2000},  glueball exchange~\cite{NakanoToki}
 or other channels with~\cite{Henley,TOY97,TOYM98} or without~\cite{Kochelev}
 suggestions
 of the hidden strangeness in the nucleon.

 One possible indication of manifestation of
 the exotic channels is non-monotonic
 behavior of the differential cross section $d\sigma/dt$ of
 $\gamma p\to \phi p$ reaction, reported recently by the
 LEPS collaboration~\cite{Mibe}. The data show a bump structure
 around $E_\gamma\simeq 2$~GeV, which disagrees  with monotonic
 behavior predicted by the conventional (Pomeron-exchange) model.
 Another peculiarity of the LEPS's data is the tendency of $d\sigma/dt$
 at forward photoproduction angle ($\theta\simeq0$)
 to be finite when the photon energy  $E_\gamma$ approaches
 to the threshold value $E_{\rm
 thr}\simeq1.574$~GeV. This is in contradiction with
 relatively old~\cite{NakanoToki,Barber} and recent~\cite{Sib06}
 expectations $d\sigma/dt=0$ at $\theta=0$ and $E_\gamma\simeq E_{\rm thr}$,
 based on a relation that near the threshold $d\sigma/dt$
 behaves as $q_\phi^2/k_\gamma^2$ where $k_\gamma$
 and $q_\phi$ are the momenta of the incoming photon and the outgoing
 $\phi$ meson in center of mass, respectively.
 The aim of present communication is to concentrate on this particular aspects
 of the experimental data. We intend to show (i) absence of
 so called "threshold factor" $q_\phi^2/k_\gamma^2$ in
 differential cross section and (ii) to stress that
 $d\sigma/dt$ at $E_\gamma\simeq E_{\rm thr}$ is sensitive to the
 dynamics of $\phi N$ interaction and is crucial for the modern  QCD inspired
 models.

\subsection{The threshold factor}
The differential cross section of $\gamma p\to \phi p$ reaction is
related to the invariant amplitude as
 \begin{eqnarray}
 {\frac{d\sigma}{dt}}^{\gamma p\to \phi p} = \frac{1}{64\pi s k_\gamma^2}
 |T^{\gamma p\to \phi p}|^2~,
 \label{2-1}
 \end{eqnarray}
where $s$ is the total energy and averaging and summing  over the
spin projections in the initial and the final states are assumed.
 The arguments lead to appearance of the threshold factor
$q_\phi^2/k_\gamma^2$ are shown in Ref.~\cite{Sib06}. First, using
the current-field identity  (vector dominance model) one can
express the invariant amplitude of the $\phi$ meson
photoproduction through the amplitudes of the
 $Vp\to\phi p$ transitions $(V=\rho,\omega,\phi)$
\begin{eqnarray}
T^{\gamma p\to \phi p}=\sum\limits_V\frac{e}{2\gamma_V} T^{Vp\to
\phi p}~, \label{1-1}
\end{eqnarray}
 where  $\gamma_\rho\div\gamma_\omega\div\gamma_\phi
\simeq 2.5\div8.5\div6.7$ are defined from the $V\to e^+e^-$
decay.
 Keeping only the diagonal transition $\phi p\to \phi p$, one can
 express the cross section of $\gamma p\to \phi p$ reaction through
 the invariant amplitude of the elastic $\phi p\to \phi p$ scattering
 \begin{eqnarray}
 {\frac{d\sigma}{dt}}^{\gamma p\to \phi p} =
 \frac{\alpha}{\gamma_\phi^2} \frac{1}{64 s k_\gamma^2}
 |T^{\phi p\to \phi p}|^2~.
 \label{2-2}
 \end{eqnarray}

 The next step is evaluating $T^{\phi p\to \phi p}(\theta=0)$.
 In~\cite{Sib06} it is made by using the optical theorem
 \begin{eqnarray}
 {\rm Im } T^{\phi p\to \phi p} (\theta=0)
 =
 -2q_\phi\sqrt{s}\sigma_{\phi p}^{\rm tot}~,
 \label{2-3}
 \end{eqnarray}
where $\sigma_{\phi p}^{\rm tot}$ is the total cross section of
$\phi p$ interaction (for convenience, we use the same sign
convention as in~\cite{Sib06}).  The consequence of
Eq.~(\ref{2-3}) is a disappearance of ${\rm Im } T^{\phi p\to \phi
p} (\theta=0)$ at $q_\phi\to 0$.  The final result reads
 \begin{eqnarray}
 {\frac{d\sigma}{dt}}^{\gamma p\to \phi p}(\theta=0)
 =
 \frac{\alpha}{16\gamma^2_\phi}
 \,\frac{q_\phi^2}{k_\gamma^2}\,
 [1 + r^2]{\sigma_{\phi p}^{\rm tot}}^2~,
 \label{2-4}
 \end{eqnarray}
where $r={\rm Re}T^{\phi p}/{\rm Im T^{\phi p}}$. Assuming $r$ to
be a constant, one can gets the threshold factor
$q_\phi^2/k_\gamma^2$ in explicit form. But the weak point of such
consideration is just assuming that $r$ is constant at $q_\phi\to
0$. The real part of invariant amplitude $T^{\phi p}$ is related
to the $\phi p$ scattering length, that can not vanish at
$q_\phi\to 0$ and therefore,
\begin{eqnarray}
r^2 (q_\phi\to 0) \sim \frac{1}{q_\phi^2}~.
 \label{2-5}
\end{eqnarray}
This leads to cancelation of $q_\phi^2$ dependence and eliminating
the "threshold factor" in Eq.~(\ref{2-4}).

\subsection{Threshold behavior of the differential cross section
}

For more consistent analysis of the threshold behavior we express
the differential cross section of $\gamma p\to \phi p$ reaction in
Eq.~(\ref{2-2}) via differential cross section of $\phi p\to \phi
p$ elastic scattering
\begin{eqnarray}
 {\frac{d\sigma}{dt}}^{\gamma p\to \phi p} =
 \frac{\alpha\pi^2}{\gamma_\phi^2k_\gamma^2}\,
{\frac{d\sigma}{d\Omega}}^{\phi p\to \phi p}~.
 \label{3-1}
 \end{eqnarray}
 At small $q_\phi$, the differential cross section
 ${d\sigma}^{\phi p\to \phi p}/{d\Omega}$
 becomes isotropic and
 it can be expressed
 through the spin averaged $\phi p$ scattering length $a_{\phi p}$
\begin{eqnarray}
{\frac{d\sigma}{d\Omega}}^{\phi p\to \phi p}=a_{\phi p}^2~.
 \label{3-1}
 \end{eqnarray}
This leads to the following estimation
\begin{eqnarray}
% {\frac{d\sigma}{dt}}^{\gamma p\to \phi p}\Big|_{\rm threshold} =
 {\frac{d\sigma}{dt}}_{\rm threshold}^{\gamma p\to \phi p} =
 \frac{\alpha\pi^2 }{\gamma_\phi^2k_\gamma^2} \,a_{\phi p}^2~.
 \label{3-2}
 \end{eqnarray}
One can see that at the threshold  the cross section of $\phi$
meson photoproduction is finite and its value is defined by the
$\phi p$ scattering length.

\subsubsection{direct estimations}

The direct estimation of the $\phi p$ scattering length on the
base of QCD sum rules was done by Koike and
Hayashigaki~\cite{KH97}. They got $a_{\phi p}\simeq -0.15$ fm
which results in
\begin{eqnarray}
 {\frac{d\sigma_{\rm thr}}{dt}}^{\gamma p\to \phi p}_{[1]} \simeq
 0.63\,{\mu\rm b }/{\rm GeV^2}~.
 \label{3-3}
 \end{eqnarray}
This value is in qualitative agreement with the experimental
indication~\cite{Mibe}.

One can estimate $a_{\phi N}$ using the $\phi N$ potential
approaches. Thus for example, Gao, Lee and Marinov suggested to
use the QCD van der Waals attractive  $\phi N$
potential~\cite{GLM} for analysis of $\phi$-nucleus bound states.
This potential reads
\begin{eqnarray}
 V_{\phi N}=-A\exp(-\mu r)/r~,
 \label{3-4}
\end{eqnarray}
 where $A=1.25$ and $\mu=0.6$~GeV. The corresponding scattering
 length  $a_{\phi p}\simeq2.37$~fm,
 found by direct solution of the Schr\"odinger equation,
 leads to large cross section
 $d\sigma/dt\simeq 1.6\times10^2\mu$b/GeV$^2$.
 It is more than two order of magnitude greater than
 the experimental hint
 and provides a problem for this potential model.
 Thus, in order to get the scattering length
 $a_{\phi p}\simeq\pm0.15$~fm (and correspondingly, the cross section
  $d\sigma/dt$ close to the experiment) one has to choose $A=2.56$ or
  0.226 for the positive (strong attraction) or negative
  (weak attraction)  $a_{\phi p}$, respectively.
  At $A\simeq 2.75$, the elastic
  scattering disappears ($a_{\phi p}=0$) and we get some kind of Ramsauer
  effect~\cite{MF}.
  In principle, such analysis may be used for other
  potentials as well.

\subsubsection{SU(3) symmetry considerations}

Estimation of the upper bound of $|a_{\phi p}|$ may be done on
assumption that the amplitudes of the $\phi p$ and $\omega p$
scattering are dominated by the scalar $\sigma$ meson exchange.
Then the SU(3) symmetry gives relation
\begin{eqnarray}
 a_{\phi p}=\xi a_{\omega p}~,
 \label{3-6}
\end{eqnarray}
where $\xi\equiv -{\rm tg}\Delta \theta_V$ ($\Delta
\theta_V\simeq3.7^0$ is the deviation of the $\phi-\omega$ mixing
angle from the ideal mixing~\cite{PDG}). More complicated
processes as $s$ channel exchange with intermediate nucleon or
nucleon resonances, or box diagrams with $\omega(\phi)\pi\rho$
vertices would give terms proportional to $\xi^2$ and generally
speaking, violate Eq.~(\ref{3-6}). But for crude estimation of
order of magnitude of $ a_{\phi p}$ one can utilize
Eq.~(\ref{3-6}) using $a_{\omega p}$ as an input.

Thus, QCD sum rule analysis of Koike and Hayashigaki~\cite{KH97}
results in $a_{\omega p}=-0.41$ fm. The coupled channel unitary
approach of Lutz, Wolf and Friman~\cite{Lutz} leads to $a_{\omega
p}=(-0.44 + \i 0.20)$ fm. An effective Lagrangian approach based
on the chiral symmetry developed by Klingl, Waas and
Weise~\cite{Klingl} results in $a_{\omega p}=(1.6 + \i 0.3)$ fm.
The corresponding $\phi$ meson photoproduction cross sections for
these scattering lengths, denoted with subscripts 2, 3 and 4,
respectively, read
\begin{eqnarray}
 {\frac{d\sigma^{\gamma p\to \phi p}_{\rm thr}}{dt}}_{[2]}
 &=&
 2.0\times10^{-2}\,{\mu\rm b }/{\rm GeV^2}~,\label{3-7}\\
 {\frac{d\sigma^{\gamma p\to \phi p}_{\rm thr}}{dt}}_{[3]}
 &=&
 2.7\times10^{-2}\,{\mu\rm b }/{\rm GeV^2}~,\label{3-8}\\
 {\frac{d\sigma^{\gamma p\to \phi p}_{\rm thr}}{dt}}_{[4]}
 &=&
 3.1\times10^{-1}\,{\mu\rm b }/{\rm GeV^2}~.
 \label{3-9}
 \end{eqnarray}

\begin{figure}[th]
 {%\centering
  \includegraphics[width=0.4\columnwidth]{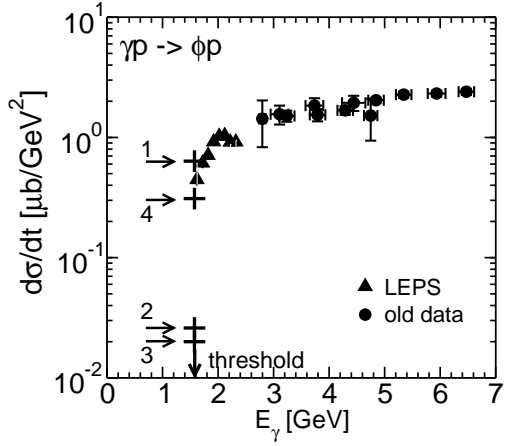}
  \caption{\label{FIG:1}{\small%\tcaps%
  Differential cross section of $\gamma p\to \phi p$ reaction
  at $\theta=0$ as function of the photon energy.
  The enumerated symbols "plus" correspond to the threshold predictions, given in
  Eqs.(\protect\ref{3-3}), (\protect\ref{3-7}) - (\protect\ref{3-9}).
  Experimental data are taken from
  Refs.~\protect\cite{Mibe,OldData}.
  }}}
 \end{figure}

  Fig.~1 shows  predictions of
  Eqs.(\ref{3-3}), (\ref{3-7}) - (\ref{3-9})
  by the enumerated symbols "plus".
  Experimental data at $\theta=0$ are taken from Refs.~\cite{Mibe,OldData}.
  The predictions of
  Eqs.~(\ref{3-3}) and (\ref{3-9}) seems to be more preferable.
  The difference between  Eqs.~(\ref{3-7}) and (\ref{3-8})
  and data can indicate small $\omega p$ scattering length or
  necessity to introduce large OZI-rule evading factor in Eq.~(\ref{3-6})
   which can be related to the finite hidden strangeness in the nucleon.
  For example, analysis of $\phi$ meson photoproduction at large
  angles in Refs.\cite{TL,Sib05} favors for the large OZI-rule
  evading factor $x_{\rm OZI}\simeq3-4$. Such value results in
  increasing the threshold predictions based on  $a_{\omega p}$ by almost
  of order of magnitude.  Employing this evading factor seems to be consistent with
  predictions $[2]$ and  $[3]$ and make a problem for that of
  $[4]$.

  \subsection{non-diagonal transitions}
\begin{figure}[th]
 {%\centering
  \includegraphics[width=0.35\columnwidth]{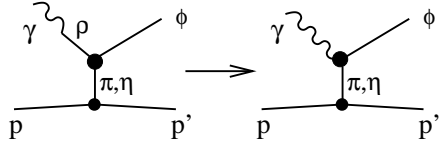}
  \caption{\label{FIG:2}{\small%\tcaps%
  Diagrammatic presentation of the pseudoscalar $\pi,\eta$ exchange
  processes in $\gamma p\to \phi p$ reaction.}}}
 \end{figure}

In principle, the non-diagonal transitions in Eq.~(\ref{1-1}) may
also contribute near the threshold. Such example is the $\phi$
meson photoproduction with $\pi (\eta)$ meson exchange , shown in
Fig.~\ref{FIG:2} , which is associated with $\rho\to \phi$
transition.

The corresponding invariant amplitude written in obvious standard
notations reads
\begin{eqnarray}
T^{\gamma p\to\phi p}_{\pi} =-i\frac{eg_{NN\pi}g_{\gamma
\phi\pi}}{M_\phi(t-m_\pi^2)}\epsilon^{\mu\nu\alpha\beta}
\varepsilon_\mu(\gamma)\varepsilon^*_\nu(\phi)
{k_\gamma}_\alpha{q_\phi}_\beta[\bar u_{p'}\gamma_5 u_p]F(t)~,
\label{4-1}
\end{eqnarray}
where $g_{\gamma\phi\pi}$ has a sense of
$g_{\rho\phi\pi}/2\gamma_\rho$ and is taking from $\phi\to
\gamma\pi$ decay ($g_{\gamma\phi\pi}\simeq0.14$~\cite{PDG}),
$g_{NN\pi}\simeq13.3$ and $F(t)$ is a product of the form factors
in $\gamma\phi\pi$ and $NN\pi$ coupling vertices. This amplitude
leads
 to the following estimate
\begin{eqnarray}
 \frac{d\sigma}{dt}^{\gamma p\to \phi p\,(\pi)}_{\rm threshold} =
  \frac{\alpha g^2_{NN\pi}g^2_{\gamma\phi\pi}F^2(t_{\rm thr})}
  {64E_{\rm thr}M_N^2M_\phi^2}
  \frac{|t_{\rm thr}|(M_\phi^2-t_{\rm thr})^2}{(t_{\rm
  thr}-m_\pi^2)^2}~,
 \label{4-2}
 \end{eqnarray}
where $E_{\rm thr} =(2M_NM_\phi+M_\phi^2)/2M_N\simeq 1.574$ GeV
and $t_{\rm thr}=-M_NM_\phi^2/(M_N+M_\phi)\simeq-0.5$~GeV$^2$.
Taking $F(t)=((\Lambda^2 -m_\pi^2)/(t-m_\pi^2))^2$ with
$\Lambda\simeq 0.6-0.7$ GeV, one can find
\begin{eqnarray}
 \frac{d\sigma}{dt}^{\gamma p\to \phi p\,(\pi)}_{\rm threshold}
 \simeq (0.8-1.6)\times 10^{-2} \mu {\rm b}/ {\rm GeV}^2~.
 \label{4-3}
 \end{eqnarray}
Coherent sum of the $\pi$ and $\eta$ meson exchange results in
 ${{d\sigma}^{\gamma p\to \phi p\,(\pi+\eta)}}/{dt}\simeq (0.3-0.6)\times 10^{-1}$
$\mu {\rm b}/ {\rm GeV}^2$.
 So again, ${d\sigma}^{\gamma p\to
\phi p}/{dt}$ is finite and its magnitude is in the range of
uncertainty of other estimations. However, being smaller than the
experimental indication it allows contribution of exotic channels
discussed in literature, such as scalar/glueball exchange, direct
knockout of hidden $\bar ss$ pairs and so on.

Finally we notice that the differential ${d\sigma}^{\gamma p\to
\phi p}/{d\Omega}$ and the total ${\sigma}^{\gamma p\to \phi p}$
cross sections have the obvious kinematical phase space factor
$q_\phi/k_\gamma$. For example, for the diagonal transition we get
\begin{eqnarray}
 {\frac{d\sigma}{d\Omega}}^{\gamma p\to \phi p}_{\rm threshold} =
 \frac{q_\phi}{k_\gamma}\frac{\alpha\pi}{\gamma_\phi^2}\,
  a_{\phi p}^2,\qquad
 {\sigma}^{\gamma p\to \phi p}_{\rm threshold} =
 \frac{q_\phi}{k_\gamma}\frac{4\alpha\pi^2}{\gamma_\phi^2}\, a_{\phi p}^2~.
 \label{4-4}
 \end{eqnarray}
If one accepts the threshold behavior of $d\sigma/dt$ as in
Eq.~(\ref{2-4}) with a constant r, then the cross sections
${d\sigma}/{d\Omega}$ and ${\sigma}^{\gamma p\to \phi p}$ will
decrease near threshold
as $(q_\phi/k_{\gamma})^3$ which seems to be rather strong.\\

In summary, we analyzed  the differential  cross section
${d\sigma}/{dt}$ of $\gamma p\to \phi p$ reaction at the threshold
and have shown that it is finite and its value is crucial for the
QCD inspired models of $\phi N$ interaction and for the mechanism
of the $\phi$ meson photoproduction.\\

 We thank E.L.~Bratkovskaya, W.~Cassing, H.~Ejiri and B.~K\"ampfer
 for useful discussions.

 %%%%%%%%%%%%%%%%%%%%%%%%%%%%%%%%%%%%%%%%%%%%%%%%%%%%%%%

\end{document}